\documentclass[10pt, twocolumn, pre, aps, superscriptaddress, showpacs]{revtex4-1}
\usepackage{amsmath, graphicx, subfigure, tikz}
\begin{document}

\title{Renormalization-Group Theory of the Heisenberg Model in d Dimensions}

\author{Egemen Tunca}
    \affiliation{TEBIP High Performers Program, Board of Higher Education of Turkey, Istanbul University, Fatih, Istanbul 34452, Turkey}
\author{A. Nihat Berker}
    \affiliation{Faculty of Engineering and Natural Sciences, Kadir Has University, Cibali, Istanbul 34083, Turkey}
    \affiliation{T\"UBITAK Research Institute for Basic Sciences, Gebze, Kocaeli 41470, Turkey}
    \affiliation{Department of Physics, Massachusetts Institute of Technology, Cambridge, Massachusetts 02139, USA}

\begin{abstract}
The classical Heisenberg model has been solved in spatial $d$ dimensins, exactly in $d=1$ and by the Migdal-Kadanoff approximation in $d>1$, by using a Fourier-Legendre expansion.  The phase transition temperatures, the energy densities, and the specific heats are calculated in arbitrary dimension $d$.  Fisher's exact result is recovered in $d=1$.  The absence of an ordered phase, conventional or algebraic (in contrast to the XY model yielding an algebraically ordered phase), is recovered in $d=2$. A conventionally ordered phase occurs at $d>2$.  This method opens the way to complex-system calculations with Heisenberg local degrees of freedom.
\end{abstract}
\maketitle

\section{Lower-Critical Dimensions and Complex Systems}

The Migdal-Kadanoff approximation \cite{Migdal,Kadanoff} has added much wide applicability to the already physically grounded position-space renormalization-group methods (Fig.1).  A very easily learned and practiced procedure, it is probably the most used renormalization-group transformation todate and today.  Among the early achievements were the nonadjustly experimentally matching global phase diagrams of surface systems \cite{BOP,krypton} starting with known microscopic potentials and the renormalization-group fixed line \cite{Jose,BerkerNelson} yielding the algebraically ordered low-temperature phase of the XY model in two dimensions. More applications, such as in a variety percolation problems \cite{percolation}, high-temperature superconductivity \cite{highTc}, ferromagnetic-antiferromagnetic \cite{Ilker2} and left-right chiral \cite{Caglar1} spin glasses, etc. followed, also quantitively obtaining the chaotic essence \cite{McKayChaos,McKayChaos2,BerkerMcKay} of the spin-glass phase.  The lower-critical dimension $d_c$ below which no ordering occurs has been correctly determined as $d_c=1$ for the Ising model \cite{Migdal,Kadanoff} and $d_c=2$ for the XY \cite{Jose} model. Most recently, the changeover from first- to second-order phase transitions of $q$-state Potts models in $d$ dimensions has been obtained by the Migdal-Kadanoff approximation.\cite{devre} In complex ordering systems with frozen microscopic disorder (quenched randomness), $d_c=2$ has been determined for the random-field Ising \cite{Machta,Falicov} and XY models \cite{Kutay}, and, yielding a non-integer value, $d_c=2.46$ for Ising spin-glass systems \cite{Atalay} (but reaching lower dimensions under spin-glass rewiring \cite{Ilker2}). Study of the Migdal-Kadanoff approximation has led to the formulation of exactly soluble hierarchical models \cite{BerkerOstlund,Kaufman1,Kaufman2}, yielding a plethora of exactly soluble models custom-fitted to the physical problems on hand.\cite{Jiang,Derevyagin2,Chio,Teplyaev,Myshlyavtsev,Derevyagin,Shrock,Monthus,Sariyer,Ruiz,Rocha-Neto,Ma,Boettcher5}

\begin{figure}[ht!]
\centering
\includegraphics[scale=0.4]{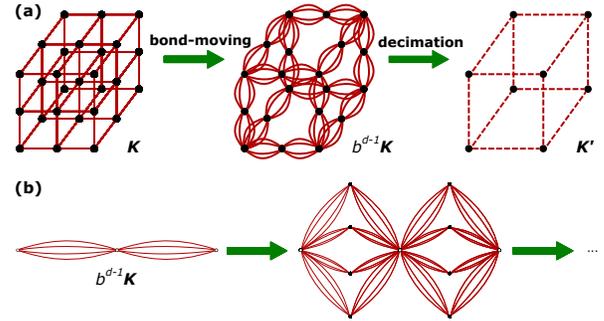}
\caption{From Ref.\cite{Artun}: (a) Migdal-Kadanoff approximate renormalization-group
transformation for the $d=3$ cubic lattice with the length-rescaling
factor of $b=2$. (b) Construction of the $d=3, b=2$ hierarchical
lattice for which the Migdal-Kadanoff recursion relation is exact. The renormalization-group solution of a hierarchical lattice
proceeds in the opposite direction of its construction.}
\end{figure}

A most important microscopic model system is the Heisenberg model, defined by the Hamiltonian
\begin{equation}
- \beta {\cal H} = J \sum_{\left<ij\right>} \, \vec s_i\cdot \vec s_j,
\end{equation}
where $\beta=1/k_{B}T$, the classical spin $\vec s_i$ is the unit spherical vector at lattice site $i$ and the sum is over all
nearest-neighbor pairs of sites.  The Heisenberg model has not been solved in the physical dimensions $d=2$ and 3 by the Migdal-Kadanoff approximation or by any other renormalization-group method.

\begin{figure}[ht!]
\centering
\includegraphics[scale=0.09]{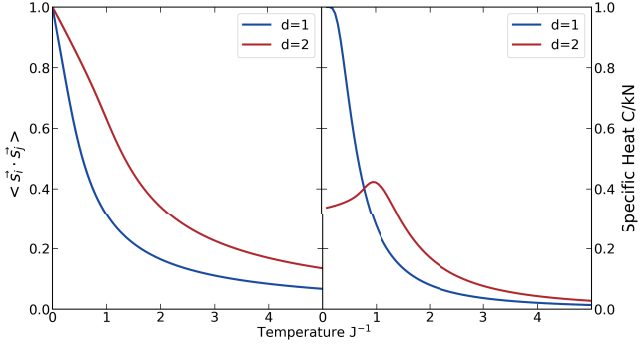}
\caption{Migdal-Kadanoff Fourier-Legendre results of the internal energy and specific heat of the Heisenberg model in $d=1,2$ from right to left in the curves. The $d=1$ curves coincide exactly with the exact calculation of Fisher \cite{Fisher}.  In $d=1,2$, there is no finite-temperature phase transition.  In $d=2$, the specific heat has a finite-temperature peak, reflecting short-range ordering \cite{BerkerNelson, Ilker2}.  As seen in this figure, no such peak occurs for the $d=1$ Heisenberg model, in contrast to the $d=1$ Ising model.}
\end{figure}

\section{Migdal-Kadanoff Renormalization Group for the Heisenberg Model}

In the first, bond-moving, step (Fig.1) of the Migdal-Kadanoff transformation,
\begin{equation}
\tilde{u}(\gamma) = u(\gamma)^2,
\end{equation}
where $u(\gamma)$ is the exponentiated nearest-neighbor Hamiltonian and $\gamma$ is the angle between the spherical unit vectors $\vec s_i$ and $\vec s_j$. The tilda denotes bond-moved. Using the Fourier-Legendre series,
\begin{equation}
u(\gamma) = \sum_{l=0}^{\infty}{\lambda_l P_l(\cos(\gamma))},
\end{equation}
with the expansion coefficient $\lambda_l$ evaluated as
\begin{equation}
\lambda_l = \frac{2l+1}{2}\int_{-1}^{1}{u(\gamma)P_{l}(\cos(\gamma))\,d(\cos(\gamma))}.
\end{equation}
Thus, for the left side of Eq.(2),
\begin{multline}
\tilde{\lambda}_{l}=\frac{2l+1}{2}\int_{-1}^{1}u(\gamma)u(\gamma)P_{l}(\cos{\gamma})\,d(\cos{\gamma}) = \frac{2l+1}{2}\,\cdot\\
\sum_{l_1=0}^{\infty}\sum_{l_2=0}^{\infty}\lambda_{l_1}\lambda_{l_2}\int_{-1}^{1}P_{l_1}(\cos{\gamma})P_{l_2}(\cos{\gamma})P_{l}(\cos{\gamma})\,d(\cos{\gamma})\\
=\sum_{l_1=0}^{\infty}\sum_{l_2=0}^{\infty}\lambda_{l_1}\lambda_{l_2}{\langle l_1 l_2 0 0 | l_1 l_2 l 0\rangle}^2,
\end{multline}
where the bracket notation is the Clebsch-Gordan coefficient with the restrictions $l_1+l_2+l=2s, s\in \mathbf{N} $; $|l_1-l_2|\leq l\leq |l_1+l_2|$.

In the second, decimation, step of the Migdal-Kadanoff transformation, a decimated bond is obtained by integrating over the shared spin of two bonds,
\begin{multline}
u'(\gamma_{13}) = \int \tilde{u}(\gamma_{12}) \tilde{u}(\gamma_{23}) \frac {d\vec s_2}{4\pi} =    \\
=\sum_{l_1=0}^{\infty}\sum_{l_2=0}^{\infty}\int \tilde{\lambda}_{l_1}\tilde{\lambda}_{l_2} P_{l_1}(\cos\gamma_{12}) P_{l_2}(\cos\gamma_{23}) \frac {d\vec s_2}{4\pi},
\end{multline}
expressing the Legendre polynomials in terms of spherical harmonics,
\begin{multline}
= \sum_{l_1=0}^{\infty}\sum_{l_2=0}^{\infty}\sum_{m_1=-l_1}^{l_1}\sum_{m_2=-l_2}^{l_2}\tilde{\lambda}_{l_1}\tilde{\lambda}_{l_2}\frac{(4\pi)^2}{(2l_1+1)(2l_2+1)}\,\cdot\\
\int Y_{l_1}^{m_1}(\vec{s_1})Y_{l_1}^{m_1*}(\vec{s_2})Y_{l_2}^{m_2}(\vec{s_2})Y_{l_2}^{m_2*}
(\vec{s_3})\frac {d\vec s_2}{4\pi},
\end{multline}
evaluating the integral and summing over the resulting delta functions,
\begin{multline}
= \sum_{l_1=0}^{\infty}\sum_{m_1=-l_1}^{l_1} \tilde{\lambda}_{l_1}^2 \frac{4\pi}{(2l_1+1)^2}Y_{l_1}^{m_1}(\vec{s_1})Y_{l_1}^{m_1*}(\vec{s_3}),
\end{multline}
due to occcuring dirac delta functions. Rearranging the spherical harmonics back to Legendre polynomials and combining with Eq.(5),
\begin{equation}
\lambda_l' = \frac{1}{(2l+1)}\big( \sum_{l_1=0}^{\infty}\sum_{l_2=0}^{\infty}\lambda_{l_1}\lambda_{l_2}{\langle l_1 l_2 0 0 | l_1 l_2 l 0\rangle}^2\big)^2,
\end{equation}
the full recursion relations of the renormalization-group are obtained.  Prime denotes renormalized.  The bond-moved $\tilde{\lambda}$ are from Eq.(5).  Thus, the renormalization-group flows are in terms of the Fourier-Legendre coefficients $\lambda_l'(\{\lambda_l\})$.  We have kept up to $l=50$ in our numerical calculations of the trajectories.

\begin{figure}[ht!]
\centering
\includegraphics[scale=0.09]{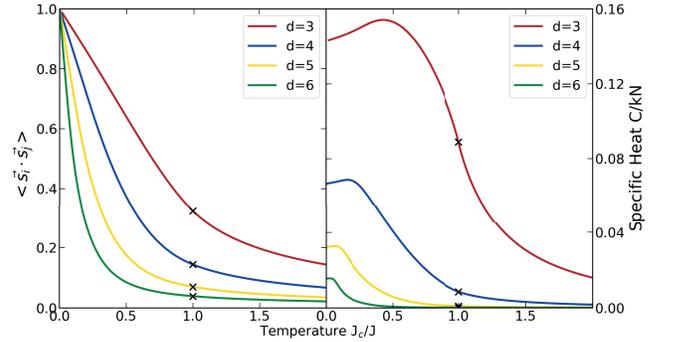}
\caption{Migdal-Kadanoff Fourier-Legendre results of the internal energy and specific heat of the Heisenberg model in $d=3,4,5,6.$ from top down in the curves.  The phase transition points are marked with $\times$.  The calculated specific heat critical exponents are $\alpha = -1.45,-2.08,-2.76,-2.92$ for $d=3,4,5,6$, respectively.}
\end{figure}

The renormalization-group trajectories are effected by repeated applications of the above transformation.  The initial points of these trajectories are obtained from the Hamiltonian in Eq.(1), which can be written as
\begin{equation}
-\beta \mathcal{H}=J\sum_{<ij>}{\vec s_i\cdot \vec s_j}=J\sum_{<ij>}{\cos{\gamma}}.
\end{equation}
Using the plane-wave expansion for the term in the partition function involving the two spins,
\begin{equation}
e^{J\cos{\gamma}}=\sum_{l=0}^{\infty}{(2l+1)i^l j_l(-iJ)P_l(\cos{\gamma})}=\sum_{l=0}^{\infty}{\lambda_l P_l(\cos{\gamma})},
\end{equation}
where $j_l(-iJ)$ is a spherical Bessel function and $P_l(\cos{\gamma})$ is a Legendre polynomial.

With no approximation, after every decimation and after setting up the initial conditions, the coefficients $\{\lambda_l\}$ are divided by the largest $\lambda_l$.  This is equivalent to subtracting a constant term from the Hamiltonian and prevents numerical overflow problems in flows inside the ordered phase.  It is also necessary in order to calculate the free energy, the internal energy, and the specific heat, as shown below.

\begin{figure}[ht!]
\centering
\includegraphics[scale=0.5]{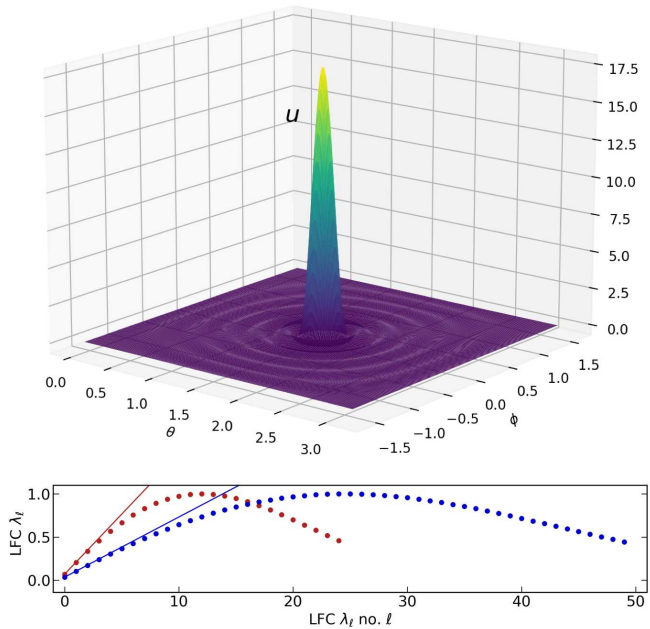}
\caption{Upper panel: Appproaching the sink (after 50 renormalization-group iterations) of the sink of the renormalization-group flows of the ordered low-temperature phase of the $d=3$ Heisenberg model. This potential function, in terms of the spherical coordinate angles $\theta$ and $\phi$ of one spin with respect to the other, is reconstructed from the renormalized Fourier-Legendre coefficients, given in the lower panel. Lower panel: Fourier-Legendre coefficients of the fixed point that is the sink of the renormalization-group flows of the ordered low-temperature phase of the $d=3$ Heisenberg model. The lines show $\lambda_l=(2l+1)\lambda_0$, valid for a delta function.  It seen that this relation is satisfied to a higher value of $l$ when a higher number of Fourier-Legendre coefficients is kept.}
\end{figure}

\section{Renormalization-Group Flows of the Fourier-Legendre Coefficients and Phase Transitions}

Under repeated applications of the renormalization-group transformation of Eq.(9), the Fouries-Legendre coefficients (FLC) flow to a stable fixed point, which is the sink of a thermodynamic phase.  The sink of the disordered phase has $\lambda_0 = 1$ and all other FLC equal to zero, $\lambda_{l>0} = 0$, meaning a constant $u$ that is not dependent on $\gamma$, namely a non-interacting system at the sink.  This sink attracts all points of the disordered phase, which it epitomizes.  In $d=1$ and $d=2$, the disordered sink is the only sink and therefore the disordered phase is the only thermodynamic phase of the system.

For $d>2$, another sink also occurs with the FLC non-zero and proportional to $2l+1$, making $u(\gamma)$ a delta function at zero separation of the spins as seen in Fig.4. (In our numerical calculation, the higher the number of kept FLC, the higher is the $2l+1$ proportionality maintained, approximating the delta function.)  This is the sink of the low-temperature ferromagnetic phase.  The disordered sink continues, as the sink of the high-temperature disordered phase.  The boundary of critical points between these two phases is controlled by an unstable fixed point, shown in Fig.5.  The largest (and only positive, since the fixed point is singly unstable) eigenvalue exponent $y$ of the derivative matrix of the recursion relations (Eq.(9)) at the critical fixed point gives the critical exponents, such the specific-heat exponent $\alpha = 2-d/y$. We calculate, at the unstable critical fixed point, $y= 0.87,0.98,1.05,1.22$ giving $\alpha=-1.45,-2.08,-2.76,-2.92$ for $d=3,4,5,6$, respectively.

\section{Renormalization-Group Calculation of Free Energy, Energy Density, and Specific Heat of the Heisenberg Model}

The Migdal-Kadanoff renormalization-group yields the entire statistical mechanics of the system, at and away from the phase transitions, including the thermodynamic quantities.  The calculation of the latter requires following the entire renormalization-group trajectories to the sink.  The logarithm of the dividing element at each operation above, namely the subtractive constant $G(n)=\ln (\lambda_{max})$, where $n$ indicates the $(n)$th renormalization-group transformation and $\lambda_{max}$ is the dividing largest FLC, summed over the trajectory, yields the free energy and therefore the thermodynamic densities.

The dimensionless free energy per bond $f = F/kN$ is thus obtained by summing the constants generated at each renormalization-group step,
\begin{equation}
f \, = \, \frac{1}{N} \ln \int_{\{s_i\}} e^{-\beta {\cal H}} \, = \,
\sum_{n=0} \frac{G^{(n)}}{b^{dn}},
\end{equation}
where $N$ is the number of bonds in the initial unrenormalized system, the first integral is over all states of the system, the second
sum is over all renormalization-group steps $n$, $G^{(0)}$ is the constant from the first division at the beginning of the trajectory.
This sum quickly converges numerically.

\begin{figure}[ht!]
\centering
\includegraphics[scale=0.5]{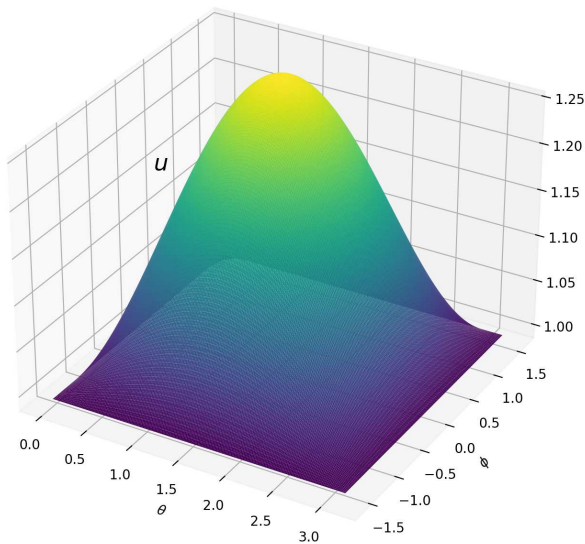}
\caption{The fixed-point potential $u(\gamma)$ of the critical point of the $d=3$ Heisenberg model. This potential function, in terms of the spherical coordinate angles $\theta$ and $\phi$ of one spin with respect to the other, is reconstructed from the renormalized coefficients of the Fourier-Legendre coefficients.}
\end{figure}

A derivative of the free energy $f$ with respect to $J$ gives the energy density $<\vec s_i\cdot \vec s_j>$.  From the dimensionless free energy per bond $f$, the specific heat $C/k N$ is calculated as
\begin{equation}
\frac{S}{k N} = f -J \frac{\partial f}{\partial J}
\end{equation}
\begin{equation}
\frac{C}{k N} = T \frac{\partial (S/k N)}{\partial T} = - J
\frac{\partial (S/k N)}{\partial J}\,.
\end{equation}

The calculated energy desities and specific heats are given in Figs.2 and 3 for $d=1,2,3,4,5,6$. The $d=1$ curves coincide exactly with the exact calculation of Fisher \cite{Fisher}.  In $d=1,2$, there is no finite-temperature phase transition.  In $d=2$, the specific heat has a finite-temperature peak, reflecting short-range ordering \cite{BerkerNelson, Ilker2}.  As seen in Fig.2, no such peak occurs for the $d=1$ Heisenberg model, in contrast to the $d=1$ Ising model.

\section{Conclusion}
We have constructed the Migdal-Kadanoff renormalization-group approximation for the Heisenberg model in $d$ dimensions and have analyzed the global renormalization-group flows, obtaining fixed points, free energies, internal energies, and specific heats.  The procedure can now be applied to Heisenberg spin glasses and Heisenberg random-field systems, and other such complex systems.

\begin{acknowledgments}
Support by the TEBIP High Performers Program of the Board of Higher Education of Turkey and by the Academy of Sciences of Turkey (T\"UBA) is gratefully acknowledged.
\end{acknowledgments}

\end{document}